\documentstyle[aps,prl,epsfig]{revtex} 
\begin{document}  
\wideabs{ 
\title{Dynamical transition from a quasi-one dimensional 
Bose-Einstein condensate to a Tonks-Girardeau gas}  

\author{P. \"Ohberg$^{1}$ and L. Santos$^{2}$}  
\address{(1) School of Physics and Astronomy, University of St. Andrews, 
North Haugh, St Andrews, Fife KY16 9SS, Scotland}
\address{(2)Institut f\"ur Theoretische Physik, Universit\"at Hannover, D-30167 Hannover,Germany}  

\maketitle  

\begin{abstract}  

We analyze in detail the expansion of a 1D Bose gas after removing the axial confinement. 
We show that during its one-dimensional expansion the density of the Bose gas does not follow a 
self-similar solution, but on the contrary, it asymptotically approaches a Tonks-Girardeau profile.
Our analysis is based on a nonlinear Schr\"odinger equation with 
variable nonlinearity whose validity is discussed for the expansion problem, by comparing with 
an exact Bose-Fermi mapping for the case of an initial Tonks-Girardeau gas. For this 
case, the gas is shown to expand self-similarly, with a different similarity law compared to the
one-dimensional Thomas-Fermi condensate.
\end{abstract}  


}  


During the last years, the achievement of Bose-Einstein condensation (BEC) \cite{BEC}
has generated an extraordinary interest in the physics of ultracold atomic gases. 
Among the topics related with the physics of ultracold gases, the issue 
of low-dimensionality is attracting a growing interest.
The recent development of trapping and cooling techniques 
has enabled experimental realizations of low-dimensional gases 
both in one \cite{MIT,Salomon,Greiner1} and two \cite{MIT,Safonov,Hansel,Florence}
dimensions. For the case of ultracold low-dimensional dilute atomic gases, 
it has been theoretically predicted that 1D \cite{Petrov1D}, 2D 
\cite{Kagan,Petrov2D} and even very elongated but still dynamically 
3D gases \cite{Petrov3D}, should present an equilibrium BEC with a spatially 
fluctuating phase. Such quasicondensates have been recently observed 
by means of time of flight measurements \cite{Hannover}. 


The impenetrable 1D gas of bosons, the so-called 
Tonks-Girardeau (TG) gas, has recently deserved a special interest 
\cite{Petrov1D,Olshanii1,Wright1,Wright3,Olshanii2}. In order to accomplish this particular 
regime, rather strict conditions for the temperature, gas density, 
interaction potential, and trapping potential must be 
fulfilled \cite{Petrov1D,Olshanii2}. These conditions can be achieved
with currently available experimental techniques. Particularly important in this sense is  
the recent progress in loading 1D Bose gases 
in optical lattices \cite{Greiner} where the transversal confinement can reach 
$100$kHz and the development of the Feshbach resonance techniques to modify the value of the 
$s$-wave scattering length \cite{Cornish}. Therefore, it is important 
to characterize the properties of the TG gas, and especially 
the intermediate regime between the quasi-1D BEC and the TG gas.


For the TG gas with a zero-range infinitely repulsive interatomic 
potential, the bosons acquire effectively a fermionic character and 
the mapping between bosonic and fermionic wavefunctions is exact, both 
for homogeneous \cite{Girardeau60} and trapped gases \cite{Wright1}.
Interestingly, the homogeneous delta-interacting 1D bosonic gas under periodic boundary 
conditions is analytically solvable for any strength of the interactions, 
as shown by Lieb and Liniger (LL) \cite{Lieb63}. 
There is unfortunately, to the best of our knowledge, no exact solution for arbitrary  
interaction strength in the case of trapped gases. 
An interesting approach was introduced in Ref.\ \cite{Kolomeisky}, where 
a hydrodynamic formalism was shown to reproduce the stationary properties of the 
TG gas. 
The approach of Ref. \cite{Kolomeisky} should, however, be employed carefully
since it significantly overestimates 
the coherence of the system \cite{Wright1}.
Recently, the approach of Ref.\ \cite{Kolomeisky} was extended to the case of 
finite interactions, by employing the LL model and local density approximation 
\cite{Olshanii2}. In Ref.\ \cite{Olshanii2} the density profile of the trapped gas was analyzed for 
regimes ranging from Thomas-Fermi (TF) profiles to TG. 
A different approach to the issue of 
finite interactions has been discussed in Ref.\ \cite{Wright2}, where the 
intermediate regime is considered as a mixture of a BEC and a fermionized TG gas.


This Letter is devoted to the analysis of the 1D expansion of Bose gas. 
We employ the procedure of Ref.\ \cite{Olshanii2} to show that contrary to the 
case of a 1D Thomas-Fermi condensate, the expansion is not self-similar. 
In fact, any initial density profile will asymptotically evolve during its expansion 
towards a TG-like shape. Consequently the 1D expansion 
allows us to easily explore intermediate situations between the TF and TG regimes.
Additionally, we discuss with the help of the Bose-Fermi (BF) mapping \cite{Wright1,Wright3}, 
the self-similar 
character of the expansion of an initial TG gas, which significantly differs from the 
self-similar expansion of a 1D TF cloud. Thus, the 1D expansion
offers a way to clearly discern between TF and TG regimes, and in between. We justify 
the validity of the employed formalism for the expansion problem by  
comparing the hydrodynamical and the BF mapping results.


We consider in the following a dilute gas of $N$ bosons 
confined in a very elongated harmonic trap with radial and axial frequencies 
$\omega_\rho$ and $\omega_z$ ($\omega_\rho\gg\omega_z$). 
If the interaction energy per particle is smaller than the 
zero-point energy $\hbar\omega_\rho$ of the transversal trap, the 
system can be considered effectively as 1D. 
We first briefly review the formalism introduced in Ref.\ 
\cite{Olshanii2}. After approximating the interparticle interaction by a 
delta function, the Hamiltonian which describes the physics of the 
1D gas becomes
\begin{equation}
\hat{H}_{\rm 1D}=\hat{H}_{\rm 1D}^{0} + 
\sum_{j=i}^{N}\frac{m\omega_{z}^{2} z_{i}^{2}}{2}
\end{equation}
with
\begin{equation}
\hat{H}_{\rm 1D}^{0}=
-\frac{\hbar^{2}}{2m}\sum_{j=1}^{N}\frac{\partial^{2}}{\partial z_{j}^{2}}
+ g_{\rm 1D}\sum_{i=1}^{N}\sum_{j=i+1}^{N}\delta \left(z_{i}-z_{j}\right)
\end{equation}
where $m$ is the atomic mass and $g_{\rm 1D}=-2\hbar^{2}/ma_{\rm 1D}$. 
The one--dimensional scattering length is 
$a_{\rm 1D}=(-a^{2}_{\rho}/2 a)[1-{\mathcal C}(a/a_{\rho})]$ \cite{Olshanii1}
with $a$ the three-dimensional scattering length, $a_\rho=\sqrt{2\hbar/m\omega_\rho}$   
the oscillator length in the radial direction, and ${\mathcal C}=1.4603\dots$.
As shown by Lieb and Liniger \cite{Lieb63}, $\hat{H}_{\rm 1D}^{0}$ 
can be diagonalized by using Bethe Ansatz \cite{Kor}. For the thermodynamic limit, 
a 1D gas at zero temperature with a given linear density $n$, 
is characterized by the energy per particle
\begin{equation}
\epsilon(n)= \frac{\hbar^{2}}{2m}n^{2} e(\gamma(n)), 
\label{energie}
\end{equation}
where $\gamma = 2/n|a_{\rm 1D}|$. The function $e(\gamma)$ fulfills
\begin{equation}
e(\gamma) = \frac{\gamma^{3}}{\lambda^{3}(\gamma)}
\int_{-1}^{1}g\left(x|\gamma\right)x^{2}dx ,
\end{equation}
where $g\left(x|\gamma\right)$ and $\lambda(\gamma)$ are the solutions of
the LL system of equations \cite{Lieb63}
\begin{eqnarray}
g\left(x|\gamma\right)&=&\frac{1}{2\pi}+\frac{1}{2\pi}\int_{-1}^{1}
\frac{2\lambda(\gamma)}{\lambda^{2}(\gamma)+(y-x)^{2}}g\left(y|\gamma\right)dy
\label{g} \\
\lambda(\gamma)&=&\gamma \int_{-1}^{1}g\left(x|\gamma\right)dx.
\label{lambda}
\end{eqnarray}
We assume next that at each point $z$ the gas is in local 
equilibrium, with local energy per particle provided by Eq.\ (\ref{energie}). Then, one can 
obtain the corresponding hydrodynamic equations for the density and the 
atomic velocity
\begin{mathletters}
\begin{eqnarray}
\frac{\partial }{\partial t}n+\frac{\partial}{\partial z} (nv)&=&0 \label{hydn}\\
\frac{\partial }{\partial t}v+v\frac{\partial}{\partial z}v&=& -\frac{1}{m}
\frac{\partial}{\partial z}(\phi(n)+\frac{1}{2}m\omega_z^2 z^2).\label{hydv}
\end{eqnarray}
\end{mathletters}
where
\begin{equation}
\phi(n) = \left (1 + n\frac{\partial}{\partial n}\right )\epsilon(n)
\label{phi}
\end{equation}
is the Gibbs free energy per particle. Inverting the corresponding Madelung transform, 
$\psi=\sqrt{n}\exp(iS)$, 
with $v=(\hbar/m)(\partial S/\partial z)$, one 
can reformulate Eqs.\ (\ref{hydn}) and (\ref{hydv}) in the 
form of a nonlinear Schr\"odinger equation (NLSE) of the form:
\begin{equation}
i\hbar\frac{\partial}{\partial t}\psi=
\left\{-\frac{\hbar^2}{2m}\frac{\partial^2}{\partial z^2}+
\frac{1}{2}m\omega_z^2z^2+\phi (|\psi|^2)\right\}\psi.
\label{NLSE}
\end{equation}
Eq.\ (\ref{NLSE}) presents similar limitations as those of the NLSE of 
Ref.\ \cite{Kolomeisky}. 
Its validity for the problem under consideration is discussed below.
Note that for the case 
of $n|a_{1D}|\rightarrow \infty$, one obtains 
$\phi(n)=g_{1D}n$, retrieving the 1D Gross-Pitaevskii equation \cite{Gross}, whereas 
for the case $n|a_{1D}|\rightarrow 0$,  one gets $\phi(n)=\pi^2\hbar^2n^2/2m$, 
and the NLSE of Ref.\ \cite{Kolomeisky} is recovered. 
The system has only one control parameter \cite{Olshanii2}, 
namely $\eta=n_{TF}^0|a_{1D}|$, where $n_{TF}^0=[(9/64)N^2|a_{1D}|/a_z^4]^3$ is the TF density, 
with $a_z=\sqrt{\hbar/m\omega_z}$. The regime $\eta\gg 1$ corresponds to the 
TF limit, in which the stationary-state density profile has a 
parabolic form. On the other hand, the regime $\eta\ll 1$ corresponds to the 
TG regime, which is characterized by a stationary-state density 
profile with the form of a square root of a parabola.


We have employed Eqs.\ (\ref{energie}),(\ref{g}),(\ref{lambda}),(\ref{phi}) and 
(\ref{NLSE}) to simulate numerically the 
expansion of a 1D gas when the axial confinement is removed, i.e. $\omega_z=0$ \cite{footnote}. 
In our simulations we have employed a Crank-Nicholson method. Special care must be paid to the 
spatial and temporal integration steps, due to the long integration times needed, the velocities 
acquired during the expansion, and the larger nonlinearity in comparison to the case of the 
standard GPE. 

\begin{figure}[ht] 
\begin{center}\ 
\psfig{file=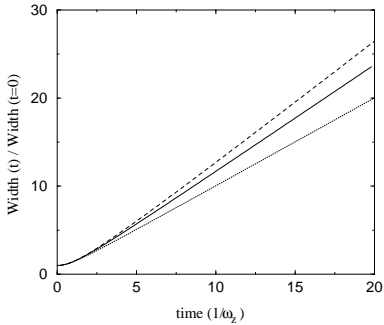,width=5.0cm}\\[0.1cm] 
\end{center} 
\caption{Cloud width $\sqrt{\langle z^2 \rangle }$ as a function of time. 
The solid line is for $\eta=1$, $\omega_r=2\pi\times 20$kHz and 
$N=200$ atoms ($\omega_z=2\pi\times 1.8$Hz at $t=0$). The dashed (dotted) lines  
are the self-similar 1D TF (TG) solutions.}
\label{fig:1}  
\end{figure}

In order to clearly understand the physics of the expansion dynamics, let us consider 
the case of a bosonic cloud which is initially TF-like ($n\sim (1-x^2/R^2)$, with 
$R$ the corresponding TF radius). This would be the case of an initial $\eta\gg 1$.
During the course of the expansion, the cloud density decreases, following 
in the first stages a self-similar TF solution 
$n(z,t)=n(z/b(t),t=0)/b(t)$ with $\ddot b=\omega_z^2/b^2$ \cite{Castin,Kagan1}. 
However, as the density decreases, the gas enters from 
the large $n|a_{1D}|$ regime into the low $n|a_{1D}|$ regime. As a consequence,  
the functional dependence of $\phi(n)$ changes throughout the whole 
cloud, and the expansion becomes no more TF self-similar (Fig.\ \ref{fig:1}). 
When this happens, the density profile departs from a parabolic TF profile, and 
asymptotically evolves towards a square root of a parabola shape, i.e. 
the density profile becomes TG-like. Similar behavior is observed for 
any initial value of $\eta$. 

\begin{figure}[ht] 
\begin{center}\ 
\psfig{file=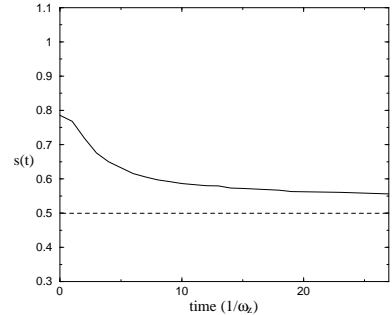,width=5.0cm}\\[0.1cm]
\end{center} 
\caption{Time evolution of the exponent $s(t)$ for $\eta=1$, for the same parameters 
as in Fig.\ \ref{fig:1}. The exponent decreases monotonically, and will asymptotically 
approach $0.5$.}
\label{fig:3}  
\end{figure}

In our simulations we have considered due to numerical limitations $\eta$ values close 
to $1$. We have analyzed in detail the dynamical transition from the initial density profile  
towards a TG shape, and observed 
that during the expansion the density profile presents at any time the form
\begin{equation}
n(z,t)=C(t)\left (1-\left (\frac{z}{R(t)}\right )^2\right )^{s(t)}
\label{app}
\end{equation}
where $R(t)$ is the radius of the cloud, and the exponent $s(t)$ takes the value $s(0)=1$ 
for an initial TF gas. The normalization constant is of the form 
$C(t)=(N/\sqrt{\pi}R(t))(\Gamma (s(t)+3/2)/\Gamma (s(t)+1))$. 
The values of $R(t)$ and $s(t)$ are determined 
at any time by means of a nonlinear least-squares fitting algorithm. In order to 
check the validity of the fit, we have also considered the normalization constant  
as a fit parameter, and compared the obtained value with the expected value $C(t)$. The
difference is less than $0.1\%$. As observed in Fig.\ \ref{fig:3} (for $\eta=1$), 
the function $s(t)$ decreases monotonically in time, and it will asymptotically reach the value $0.5$
The function $s(t)$ presents two clear time scales. It decreases fast during the first stages 
of the expansion (few axial trap periods), but the final convergence towards the TG profile 
is significantly slower. The latter 
is expected since the exact $n^2$ dependence of the functional $\phi(n)$ 
just appears asymptotically.  
We have analyzed the expansion for larger initial values of $\eta$, and observed a similar behavior, 
although for larger values of $\eta$ the convergence towards $s=0.5$ requires longer times (for $\eta=3$ it 
should occur at around $350$ axial trap periods).
For practical purposes, if $\eta$ is not sufficiently close to 1, only the first stage of the 
evolution of $s(t)$ will be observable, since for longer time scales the density will significantly 
decrease. For $\eta\gg 1$, the usual TF self-similar solution ($s(t)=const=1$) is retrieved 
for any practical purposes.
Note, however, that the transition into a TG profile during a 1D free expansion is in principle 
unavoidable for whatever initial condition, i.e. the $s$ coefficient tends to $0.5$. This is supported 
by the fact that the density profile fulfills at any time Eq.\ (\ref{app}), and that 
the only self-similar solution supported by the NLSE (\ref{NLSE}) at very low densities is 
with $s=0.5$. 


Once shown that Eq.\ (\ref{NLSE}) predicts that during a free 1D 
expansion the density profile dynamically becomes TG-like, 
let us analyze the validity of the equation for the problem 
under consideration. As shown in Ref.\ 
\cite{Wright1}, the hydrodynamical approach should be carefully employed, since 
it overestimates the coherence in the system. In order to check that Eq.\ (\ref{NLSE}) 
provides the right physical picture in our problem, 
we have calculated the free expansion of an initial TG gas using 
both the BF map, and the NLSE. 

From the BF map one obtains that the dynamics of 
the density profile for an impenetrable gas of bosons is given by \cite{Wright1} 
\begin{equation}
n(z,t)=\sum_{n=0}^N |\phi_n(z,t)|^2,
\label{BFmap}
\end{equation}
where $\phi_n(z,t)$ denotes the time-dependent wavefunction of the 
$n$-th eigenmode of the original axial harmonic oscillator.
The expansion dynamics for each $\phi_n$ is obtained analytically 
by means of the corresponding Green function in free space 
\begin{equation}
G(z-z',t)=-i\left ( \frac{m}{2\pi\hbar it} \right )^{1/2}
e^{im(z-z')^2/2\hbar t}.
\label{Green}
\end{equation}
From Eq.\ (\ref{BFmap}) one obtains a self-similar solution of the form
\begin{equation}
n(z,t)=\frac{1}{\sqrt{1+\omega_z^2 t^2}}
n\left (\frac{z}{\sqrt{1+\omega_z^2 t^2}},t=0\right ).
\label{selfsimilar}
\end{equation}
Note, that for times $t\gg 1/\omega_z$ 
the scaling coefficient $\sqrt{1+\omega_z^2 t^2}$ becomes $\omega_z t$, 
whereas for the case of a 1D TF self-similar solution 
the scaling coefficient becomes $\sqrt{2}\omega_z t$. 
Consequently, the expansion of an initial TG and TF gas is significantly different. 
This property could be employed to discern between the two regimes.

From the corresponding hydrodynamic equations (\ref{hydn}) and (\ref{hydv}), 
one can easily prove that the same self-similar solution (\ref{selfsimilar}) 
for the density 
is obtained from Eq.\ (\ref{NLSE}) in the limit of $n|a_{1D}|\rightarrow 0$, 
i.e. using the equation of Ref.\ \cite{Kolomeisky}. 
Therefore, Eq.\ (\ref{NLSE})
accurately describes the expansion dynamics even for the extreme case 
of a TG gas, and it is thus expected to describe well 
the expansion for intermediate regimes between the TG and 
the TF limits, where the coherence is not yet completely lost.


In this Letter, we have studied the dynamical transition from a quasi-1D BEC into a TG gas 
during the expansion. Our analysis is based 
on a NLSE with variable nonlinearity, which generalizes for arbitrary 
interaction the extremal cases provided by the Gross-Pitaevskii equation 
($n|a_{1D}|\rightarrow\infty$) and the equation of Ref.\ \cite{Kolomeisky}
($n|a_{1D}|\rightarrow 0$). We have shown that even if the initial cloud possesses a TF profile, 
the density profile acquires asymptotically a TG shape. We have analyzed in detail this 
transition, and characterized the shape of the cloud in the intermediate stages.
We have evaluated by means of a BF map the exact expansion dynamics of a 
TG gas and shown that the expansion is self-similar with a significantly different scaling law 
compared to a TF gas. We have additionally shown that the NLSE approach provides exactly 
the same self-similar solution as the BF map for the case of a TG gas, and it is therefore 
expected to describe well the expansion for any intermediate regime.

Let us additionally point out that the NLSE (\ref{NLSE}) also provides 
the excitation spectrum of the 
1D Bose gas in intermediate regimes between TF and TG, by considering 
a small perturbation around the ground state solution $\psi_0(z)$
of Eq. (\ref{NLSE}) $\psi(z)=\psi_0(z)+\delta\psi(z)$, where $\delta\psi$ is given by 
$\delta\psi(z)=u(z)e^{-i\omega t}+v(z)^*e^{i\omega t}$. Inserting this Ansatz into 
Eq. (\ref{NLSE}) leads to the corresponding Bogoliubov-de Gennes equations 
\begin{eqnarray}
{\cal L}u(z)+n_0\phi'(n_0)v(z)&=&\hbar\omega u(z)  \label{bdg1}\\
-{\cal L}v(z)-n_0\phi'(n_0)u(z)&=&\hbar\omega v(z) \label{bdg2}
\end{eqnarray} 
where $n_0=\psi_0^2$, $\phi'=d\phi/dn$, and 
$ {\cal L}=-(\hbar^2/2m)(\partial^2/\partial z^2)+m\omega_z^2 z^2/2+
\phi(n_0)+n_0\phi'(n_0)-\mu$, with 
$\mu$ the chemical potential fixed by the normalization of $n_0$.
Eqs.\ (\ref{bdg1}) and (\ref{bdg2}) describe the crossover from the TF to the TG
regime for all excitation frequencies. 
In particular, we have obtained that these equations provide the same results as
in Ref.\ \cite{Chiara} for the lowest compressional mode. 

Summarizing, the 1D expansion dynamics constitutes an 
experimentally accessible tool to discern between the different interaction regimes 
in a 1D gas, and additionally provides a way to dynamically accomplish the TG gas. 
Unfortunately,  the method employed in this Letter does not allow to analyze the 
fundamental problem of decoherence when entering the TG regime. 
The solution of this problem requires to extend the exact results of Refs.\ \cite{Wright1,Wright3} 
to the case of inhomogeneous time dependent Bose gases with finite interactions, 
in which the BF mapping is not exact. Such analysis is beyond the scope of this Letter 
and it will be the subject of future investigations.


We acknowledge  support from Deutsche Forschungsgemeinschaft (SFB 407),  
the RTN Cold Quantum gases, ESF PESC BEC2000+ and EPSRC. 
L. S. wishes to thank the Alexander von Humboldt Foundation, 
the Federal Ministry of Education and 
Research and the ZIP Programme of the German Government.
Discussions with M. Lewenstein, D. S. Petrov, and G. V. Shlyapnikov are acknowledged.


\begin{references}
\bibitem{footnote} We have calculated the function $\phi$ directly from the LL equations, and compared our 
results with those provided in Ref. [19] of Ref. \cite{Olshanii2}, obtaining an excellent agreement.
\end{references}
\end{document}